\documentclass[preprintnumbers,showpacs,amsmath,amssymb,floatfix,prd,onecolumn,superscriptaddress,nofootinbib]{revtex4}
\usepackage{graphicx}
\usepackage{epsfig}
\usepackage{bm}
\usepackage{amsfonts}
\usepackage{epstopdf}
\usepackage{bm}
\usepackage{amsthm}
\usepackage{multirow}
\usepackage{subfigure}
\usepackage{verbatim}

\begin{document}

\title{Dynamics of Viscous Phantom Universe}

\author{Ji-Yao Wang}
\email{wjykana@foxmail.com}
\affiliation{Division of Mathematical and Theoretical Physics, Shanghai Normal University, 100 Guilin Road, Shanghai 200234,  P.R.China}

\author{Chao-Jun Feng}
\email{fengcj@shnu.edu.cn}
\affiliation{Division of Mathematical and Theoretical Physics, Shanghai Normal University, 100 Guilin Road, Shanghai 200234,  P.R.China}

\author{Xiang-Hua Zhai}
\email{zhaixh@shnu.edu.cn}
\affiliation{Division of Mathematical and Theoretical Physics, Shanghai Normal University, 100 Guilin Road, Shanghai 200234,  P.R.China}

\author{Xin-Zhou Li}
\email{kychz@shnu.edu.cn}
\affiliation{Division of Mathematical and Theoretical Physics, Shanghai Normal University, 100 Guilin Road, Shanghai 200234,  P.R.China}

\begin{abstract}
The phantom dark energy remarkably boosts our prehension of the accelerating Universe. Various models are widely discussed  in the phantom Universe without bulk viscosity. From the hydrodynamics' point of view, it is natural to introduce the nonperfect fluid in the study of the Universe, as an ideal fluid just an  approximation to the real world after all and using the generalized equation of state (EoS) with bulk viscosity, the  early inflationary universe and the accelerated expansion of the late-time universe are described by many authors. In this paper, in order to  investigate how the viscosity will influence the evolution of the Universe, we study a class of phantom dark energy models with bulk viscosity by the method of dynamical analysis technique.  We show that there are different cosmic late-time behaviors and the stability also brings some constraints on the models. We also plot the evolutionary trajectories of this model in the statefinder parameter-planes to see the different behaviors of the models from the statefinder viewpoint. 
\end{abstract}

\maketitle


\maketitle


\section{Introduction}
In the past 20 years, the cosmological observational data made it possible to understand the geometry and the expansion history of the Universe \cite{Riess:1998cb,Perlmutter:1998np,Spergel:2003cb,Eisenstein:2005su,Kowalski:2008ez,Aghanim:2018eyx,Hinshaw:2012aka}. The late time acceleration of the Universe indicates the requirement of either the modification of theories of gravitation or the existence of a component in the Universe that acts as gravitational repulsion. Such a component, which should be relatively uniform in the observable Universe, is called  dark energy, whose  origin is still an open question in modern physics. Dark energy is usually described by the equation of state (EoS) $w=p/\rho$, where $p$ is the pressure and $\rho$ is the density.  From the evolutionary equation of the scale factor $a$:
\begin{equation}
\frac{\ddot{a}}{a}=-\frac12(1+3w)\frac{\dot a^2}{a^2}\,,
\end{equation}
we can see that an accelerating Universe desires $w<-\frac{1}{3}$, which suggests that the pressure of dark energy is negative. Here we use the units that $8\pi G=1$. With the equation of state  $w=-1$, the cosmological constant $\Lambda$ has provided a good description for the accelerating Universe. However, one cannot explain why the observed cosmological constant is so small when  the constant is considered as quantum vacuum energy, which is one of the most popular explanation of the cosmological constant. The value of  vacuum energy density contributed from the sum of  all vacuum modes below an ultraviolet cut-off at the Planck scale is given by $\rho_\Lambda\sim10^{112} {\rm erg/cm}^3\,$, which  exceeds the observational value of $\rho_\Lambda\sim10^{-8} {\rm erg/cm}^3$  by about 120 orders of magnitude\cite{Weinberg:2000yb}.

Therefore, different dark energy models other than the cosmological constant have been suggested to describe the accelerating Universe. Among the models, the scalar field models of dark energy  may probe the nature of the acceleration of the Universe. Such a kind of canonic scalar field is called quintessence\cite{Peebles:1987ek,Ratra:1987rm}, which is considered as one of the candidates of dark energy that inspired by the quantum theory. One the other hand, the cosmological observational data shows that the EoS $w_{DE}$ for dark energy lies in a narrow range near the value $w=-1$. So R.Caldwell suggested another type of scalar  field in the Universe named phantom with $ w<-1$\cite{Caldwell:1999ew,Caldwell:2003vq},  which differs from the canonic action for the scalar field only by the sign of the kinetic term.  In Ref.\cite{Li:2005ay,Hao:2003ww,Hao:2003th,Li:2003ft}, the authors pointed out that the big rip in late Universe of phantom dark energy can be avoided. More generalized dynamical models of dark energy like quintom \cite{Guo:2004fq}, k-essence\cite{Rendall:2005fv} and H-essence\cite{Wei:2005nw} are also widely discussed. 

Cosmology with viscosity is also an interesting alternative to understand the expansion of the Universe. From the hydrodynamics' point of view, it is natural to introduce the nonperfect fluid in the study of the Universe, as an ideal fluid just an  approximation to the real world after all.  The evolution of nonperfect fluid is a dissipative process, which can be described by bulk viscosity, shear viscosity, and heat conduction. The viscous relativistic fluids were first suggested in Refs.\cite{Eckart:1940te,landau}. In the general theory of dissipation in relativistic nonperfect fluid, the evolution equation becomes very complicated. Fortunately, if we study the phenomenon in the quasi-thermal equilibrium state, the conventional
theory is still valid. In fact, in the homogeneous and isotropic Universe, the dissipative processes can  be described by bulk viscosity and the shear viscosity can be ignored. The bulk viscosity introduces dissipation by re-defining the effective pressure $p_{eff}$ as 
\begin{equation}
p_{eff}=p_i-3\xi_i H\,,
\end{equation}
where $\xi_i$ is the bulk viscosity coefficient of any component and
$H$ is the Hubble parameter. 

The interest in viscous universe has increased in recent years\cite{Zhai:2005mu,Brevik:2006md,Capozziello:2005pa,Nojiri:2005sr,Brevik:2011mm,Brevik:2017msy,Normann:2016jns,Sun:2009pb,Feng:2009jr,Hu:2005fu,Koivisto:2005mm,Meng:2005jy,Pourhassan:2013sw,Hernandez-Almada:2020ulm,Nojiri:2006zh}. For example, in Ref.\cite{Zhai:2005mu},   the authors studied the cosmological dynamics of the viscous generalized Chaplygin gas , giving the constraints of the parameters. In Ref.\cite{Sun:2009pb}, the authors discussed the viscous Cardassian models and fit the models with Ia SN data, which is instructive for the study of observational cosmology. The authors of Ref.\cite{Feng:2009jr}  alleviated the cosmological age problem by  investigating the viscous Ricci dark energy. And the authors of Ref.\cite{Hernandez-Almada:2020ulm} did a statistics analysis considering an interacting and viscous Universe
and  performed a dynamics system approach as well.

On the other hand, the dynamical system of the Universe is a non-linear system so it is hard to  find its analytic solution. In order to describe the evolution of a dynamical system, people usually find the critical points of the system and study the perturbations around these critical points to determine the stability of the system. Another advantage of the dynamical approach is that it can completely avoid the influence of non-linear effect. Thus, in cosmology, the dynamical method is widely used in the discussions of dark energy\cite{Feng:2012wx} or modified gravitation\cite{Feng:2014fsa}. 
General reviews of the autonomous systems in Friedman-Lemaitre-Robertson-Walker(FLRW) Universe have been given in Refs.\cite{Bahamonde:2017ize,Copeland:2006wr}.

In this paper, we,will assume the existence of the bulk viscosity in the phantom dark energy and investigate the evolution of the viscous phantom Universe by the dynamical approach. The paper is organized as follows. In Sec.\ref{sec:model} we give a brief review of the phantom dark energy and reconstruct the viscous phantom models. We study three different models A, B and C of viscous phantom dark energy in Sec.\ref{sec:m1}, Sec.\ref{sec:m2} and Sec.\ref{sec:m3}. Among the three models, Model C is a special case with tracking attractor. And in Sec.\ref{sec:sts}, we apply statefinder diagnostic to differentiate among different forms of phantom Universe.  Finally, discussions and conclusions will be given in Sec.\ref{sec:conclusion}.

\section{Viscous Phantom Dark Energy}\label{sec:model}

The FLRW metric that describes a homogeneous and isotropic flat Universe is given by
\begin{equation}
ds^2 = -dt^2 + a(t)^2\bigg[dr^2+r^2(d\theta^2 + \sin^2\theta d\phi^2)\bigg]\,,
\end{equation}
where $a(t)$ is the scale factor. 

The action for the phantom field  minimally coupled to gravity is given by
\begin{eqnarray}
S=\int d^4x\sqrt{-g}\left[-(\partial \phi)^2+V(\phi) \right]\,.
\end{eqnarray}
The Friedmann equations of the Universe composed by dust matter and phantom field read
\begin{eqnarray}
H^2&=&\frac{1}{3}\left( \rho_m+\rho_\phi \right)\,,\\
\dot{H}&=&-\frac{1}{2}\left( \rho_m+\rho_\phi+p_\phi \right)\,,
\end{eqnarray}
where $\rho_i$'s and $p_i$'s are the densities and pressures of different components. 

The dissipation of bulk viscosity is introduced by the effective pressure   of phantom field\cite{Eckart:1940te,Zhai:2005mu,Sun:2009pb,Feng:2009jr}
\begin{eqnarray}
p_{eff}=p_\phi-3\xi_\phi H\,,
\end{eqnarray}
and the evolutionary equation of phantom field can be written as
\begin{eqnarray}
\ddot{\phi}+3H\dot{\phi}-\frac{dV(\phi)}{d\phi}+\frac{9\xi_\phi H^2}{\dot{\phi}}=0\,,
\end{eqnarray}
where $\xi_\phi$ is the bulk viscosity coefficient of phantom field. Due to the second law of thermodynamics, we have $\xi_\phi>0$ that assures a positive entropy production.  So the evolution equations for the dust matter and the phantom fields can be written as
\begin{eqnarray}
\dot{\rho}_m+3H\rho_m&=&0\,,\\
\dot{\rho}_\phi+3H(\rho_\phi+p_\phi-3H\xi_\phi)&=&0\,.
\end{eqnarray}

We introduce the dimensionless variables as follows:
\begin{eqnarray}
x=\frac{\dot{\phi}}{\sqrt{6}H}\,\,&,&\,\,y=\frac{\sqrt{V(\phi)}}{\sqrt{3}H}\,,\\\nonumber
\lambda=-\frac{V^\prime(\phi)}{V(\phi)}\,\,&,&\,\,\Gamma=\frac{V(\phi)V^{\prime\prime}(\phi)}{V^\prime(\phi)^2}\,\,\,,\,\,\zeta=\frac{\xi_\phi}{3H}\,\,.\\\nonumber
\end{eqnarray}

The relative densities of the components are given by
\begin{eqnarray}
\Omega_m&=&1+x^2-y^2\,,\\
\Omega_\phi&=&-x^2+y^2\,.
\end{eqnarray}
And the EoS of phantom field reads
\begin{eqnarray}\label{eqs:w}
w_\phi=\frac{p_\phi}{\rho_\phi}=\frac{-x^2-y^2-\frac{\xi_\phi}{H}}{-x^2+y^2}\,.
\end{eqnarray} 
The total EoS parameter in terms of $H$ reads
\begin{eqnarray}\label{eqs:hubb}
w_{tot} = \frac{p}{\rho }=- 1 - \frac{2}{3}\frac{\dot H}{H^2}\,,
\end{eqnarray}
where
\begin{eqnarray}
\dot{H}/H^2=-\frac{3}{2}\left( 1-x^2-y^2+\frac{\xi_\phi}{3H}  \right)\,.
\end{eqnarray}
For viscous phantom cosmological dynamical system, the equations of autonomous system can be expressed as
\begin{eqnarray}
\frac{dx}{dN}&=&-3x-\frac{\sqrt{6}}{2}\lambda y^2+\frac{3}{2}x\bigg[1-x^2-y^2-\zeta\bigg]-3\zeta\,,\\
\frac{dy}{dN}&=&-\frac{\sqrt{6}}{2}\lambda xy+\frac{3}{2}y\bigg[1-x^2-y^2-\zeta \bigg]\,,\\
\frac{d\lambda}{dN}&=&-\sqrt{6}\lambda^2x(\Gamma-1)\,.\\
\end{eqnarray}
where $N=\ln a=-\ln (1+z)$. We will study three  models with different viscosity in the following sections.

\section{Autonomous system of  Model A: $\xi_\phi=3\xi_0 H$}\label{sec:m1}

Firstly we are interested in the model whose bulk viscosity is proportional
to the Hubble parameter, given by $\xi_\phi=3\xi_0 H$. In this model, we choose the exponential potential as
\begin{eqnarray}
V(\phi)=V_0 e^{-\alpha \phi}\,,
\end{eqnarray}
where $\alpha$ and $V_0$ are positive constants. Thus, the  equations of dynamical system of Model A can be reduced as the following equations:
\begin{eqnarray}
\frac{dx}{dN}&=&-3x-\frac{\sqrt{6}}{2}\alpha y^2+\frac{3}{2}x\bigg[1-x^2-y^2-\xi_0\bigg]-3\xi_0\,,\\
\frac{dy}{dN}&=&-\frac{\sqrt{6}}{2}\alpha xy+\frac{3}{2}y\bigg[1-x^2-y^2-\xi_0 \bigg]\,,
\end{eqnarray}
from which the critical points are obtained and the physical conditions of these critical points can be studied, see Table I. And we may also perform perturbations to study the stability of these points by substituting linear perturbations near the critical points in the form as
\begin{eqnarray}
x&=&x_{(Ai)}+\delta x\,,\\
y&=&y_{(Ai)}+\delta y\,,\,i\in (1,2)\,,
\end{eqnarray}
where $x_{(Ai)}$ and $y_{(Ai)}$ denote the coordinates of the critical points $(x_{(Ai)}\,,\,y_{(Ai)})$. From the perturbation equations
\begin{eqnarray}
\frac{d}{dN}\left(                 
\begin{array}{c}   
 \delta x  \\ 
\\ 
 \delta y  \\  
\end{array}
\right)=\left(                 
\begin{array}{cc}   
-\frac{3}{2}\big(1+3x^2+y^2+\xi_0\big) &-3xy-\sqrt{6}\alpha y \\
\\ 
-3xy-\sqrt{\frac{3}{2}}\alpha y &\frac{1}{2}\big(3-3x^2-9y^2-\sqrt{6}\alpha x-\xi_0\big) \\  
\end{array}
\right)\left(                 
\begin{array}{c}   
 \delta x  \\ 
\\ 
 \delta y  \\  
\end{array}
\right)\,,
\end{eqnarray}
we may give the eigenvalues of each point. For a 2D autonomous system, a stable point requires the real parts of both eigenvalues to be negative, a saddle point is with one eigenvalue having a real part and the other having a negative one, and the point is unstable if the real parts of both eigenvalues are positive. Ignoring the unphysical points, we list the two fixed points of the autonomous equations in Table I, labelled as Case (A1) and Case (A2).
\begin{table}[h]
	\centering
	\label{tab:m1}
	\begin{tabular}{cccc}
		\hline
		\hline
		Case &$\,\,\,\,$  Critical points $(x,y)$$\,\,\,\,$   &Parameter Region (Fig.1) &Stability \\ \hline
		\\
		&&I &Stable\\
		\\
		(A1)&$\big(x_{(A1)}\,,\,0\big)$&II$\&$III &Saddle\\
        \\
        &&IV &Unstable
        \\
        \\
        \hline
        \\
		\multirow{4}{*}{(A2)} &\multirow{4}{*}{$\big(x_{(A2)}\,,\,y_{(A2)}\big)$}&III$\&$IV&Stable\\
		\\
		&&I$\&$II &Saddle\\	
		\\	
		\hline
		\\
		\multicolumn{4}{c}{{ $x_{(A1)}=-\frac{1+\xi_0}{3^{1/3}}\Lambda_A+(9^{1/3}\Lambda_A)^{-1}\,\,,\,\,		\Lambda_A=\big(-9\xi_0+\sqrt{3}\sqrt{\xi_0^3+30\xi_0^2+3\xi_0+1}   \big)^{-1/3} $\,.}}\\
	\\
	\hline
	\\
			\multicolumn{4}{c}{{ $		x_{(A2)}=\frac{-\alpha^2+3-\Delta_A}{2\sqrt{6}\alpha}\,\,,\,\,y_{(A2)}=\sqrt{1-\frac{\sqrt{6}}{3}\alpha x_{(A2)}-\xi_0-x_{(A2)}^2}$\,,\, $\Delta_A=\sqrt{\alpha^4+6\alpha^2-12\alpha^2\xi_0+12\sqrt{6}\alpha\xi_0+9}$\,.}}\\
		\hline
		\hline
	
	\end{tabular}

	\caption{The fixed points and their stability of autonomous system of Model A. The detail of the physical conditions are shown in Appendix A.}
\end{table}

For Case (A1), the eigenvalues of the perturbation matrix are given by
\begin{eqnarray}
\lambda_1&=&-\frac{3}{2}\big(1+3x_{A1}^2+\xi_0\big)\,,\\
\lambda_2&=&\frac{1}{2}\big(3-3x_{A1}^2-\sqrt{6}\alpha x_{A1}-\xi_0\big)\,.
\end{eqnarray}
It is conceivable that the stability of this critical point is determined by the values of $\alpha$ and $\xi_0$. The corresponding regions of the two parameters to the stability are shown in Fig.1(a) and  also listed in Table I. This point trends to $(0,0)$ when $\xi_0\rightarrow 0$, thus it represents a matter dominated Universe.

Case (A2) is the only non-trivial critical point besides case (A1), which represents a phantom dominated Universe.  The eigenvalues of the perturbation matrix are given by
\begin{eqnarray}
\lambda_3&=&\frac{1}{4}\bigg[C_1 x_{A2}+C_2 -\sqrt{D_1x_{A2}^3+D_2x_{A2}^2+D_3x_{A2}+D_4}  \bigg]  \,,\\
\lambda_4&=&\frac{1}{4}\bigg[C_1 x_{A2}+C_2  +\sqrt{D_1x_{A2}^3+D_2x_{A2}^2+D_3x_{A2}+D_4} \bigg]  \,,
\end{eqnarray}
where $C_1$, $C_2$ and $D_1$ to $D_4$ are  coefficients composed of $\alpha$ and
$\xi_0$ (see Appendix A). The regions of $\alpha$ and $\xi_0$ indicating the stability of the critical point $(A_2)$  are illustrated in Fig.1(b) and the stable areas of (A1) and (A2) are separate. However, in a physical view, we want $\alpha$ and $\xi_0$  to be small. So in the following study we may consider case (A1) as a saddle point and case (A2) as a stable point.

Next, we study the above dynamical system numerically. We choose the parameter to be $\alpha=0.2$ and $\xi_0=0.01$. The  phase diagram is  given in Fig.2(a), where we can observe that all orbits tend to an attractor which describes a late time transition from matter dominated era to phantom dominated one.  Note that as we increase of the value of $\alpha$ and $\xi_0$ individually, the attractor moves away from the point $(0,1)$ in  different directions, which are shown  in Fig.2(b) and Fig.2(c). These results drop a hint that the viscosity does play an important role in the evolution of the
Universe.

\begin{figure}
	\centering
	\includegraphics[width=0.3\linewidth]{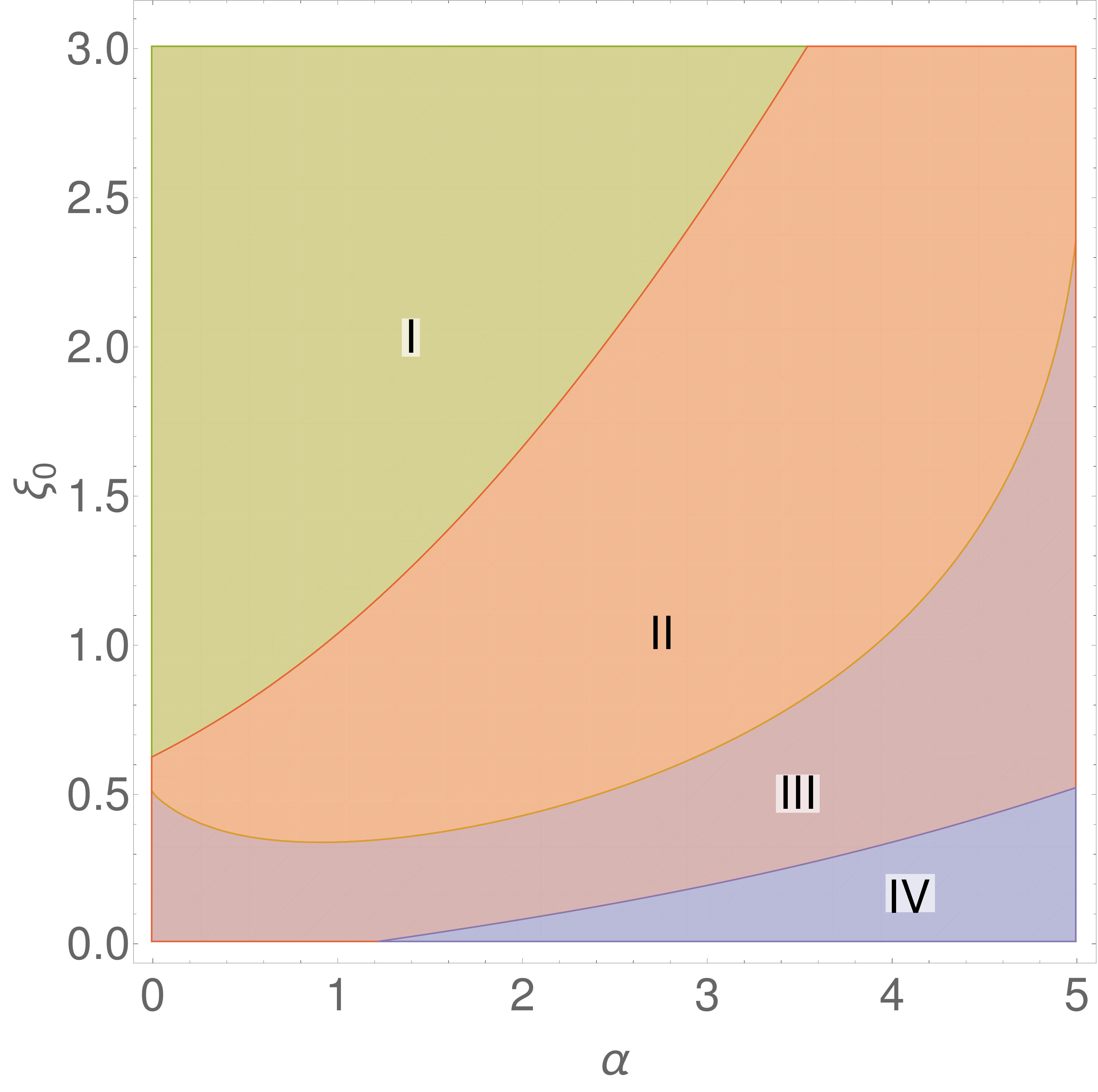}
	\caption{The stable region of Case (A1) and Case (A2). In Figure.(a), the yellow region named "I" is the stable region of Case (A1). Similarly the green region named "II" is saddle region and the red one named "III" is unstable region. The stable condition of Case (A2) is shown in Figure.(b), the yellow region named "I" is the stable region and green region named "II" is saddle region. In Figure.(c), the upper pink region is the stable region for Case (A1) and the lower yellow region is the stable region for Case (A2). It is clearly shown that the stable region of the two fixed points separate from each other.}
	\label{fig:bounda1}
\end{figure}

\begin{figure}[htbp]
	\centering
	\subfigure[The phase graph of Model A $\{\alpha\,,\,\xi_0\}=\{0.2\,,\,0.01\}$.]{
		\includegraphics[width=0.25\linewidth]{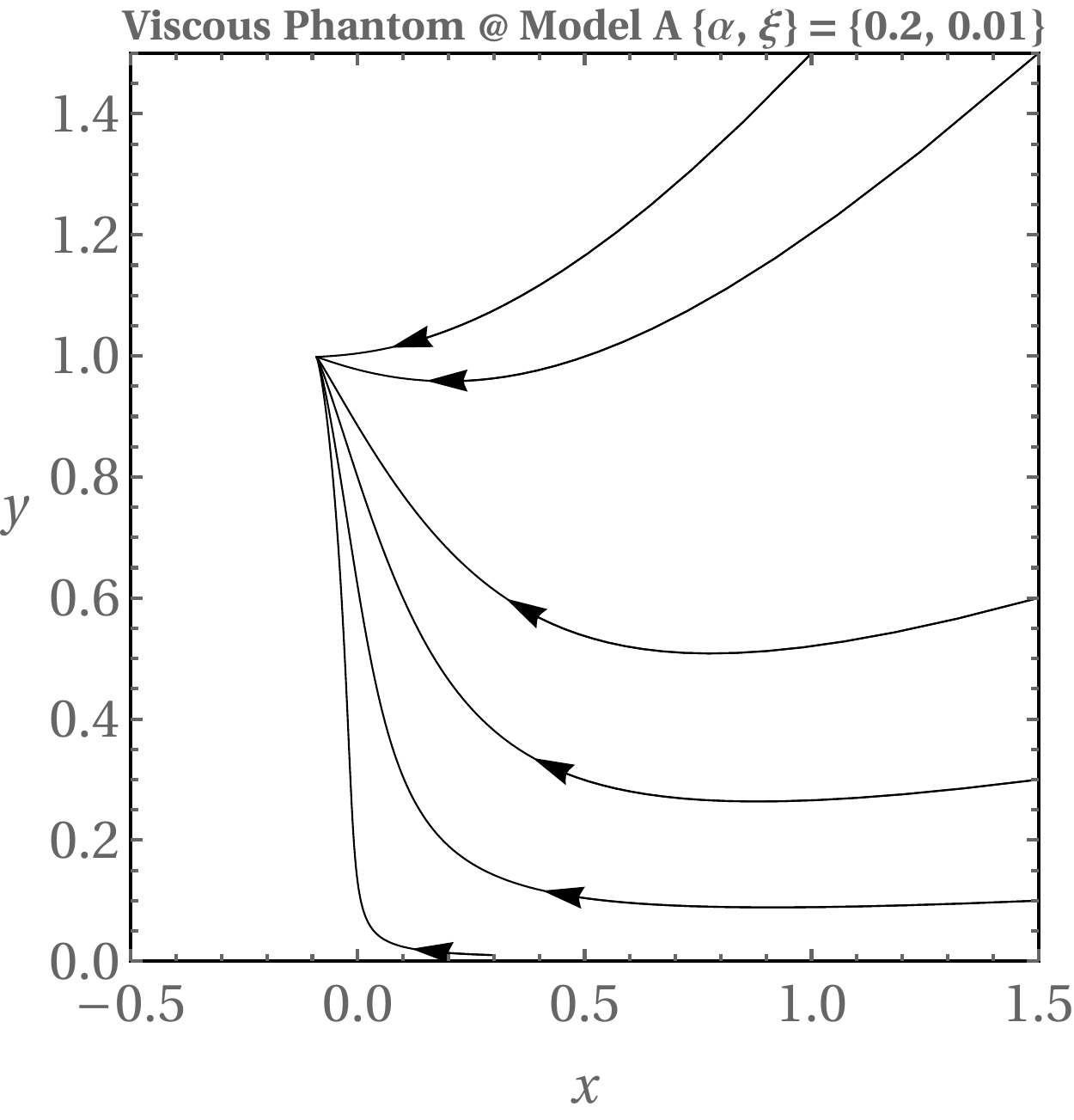}
	}
\quad
	\subfigure[The variation of the node point $(x_{(A2)}\,,\,y_{(A2)})$, $\xi_0$ is fixed as $\xi_0=0.01$ and $\alpha$ varies from 0 to 1. The critical points when $\{\alpha,\xi_0\} =\{0.1,0.01\}$ and $\{\alpha,\xi_0 \} =\{0.2,0.01 \}$ are specially mentioned on the curve.]
	{
	\includegraphics[width=0.25\linewidth]{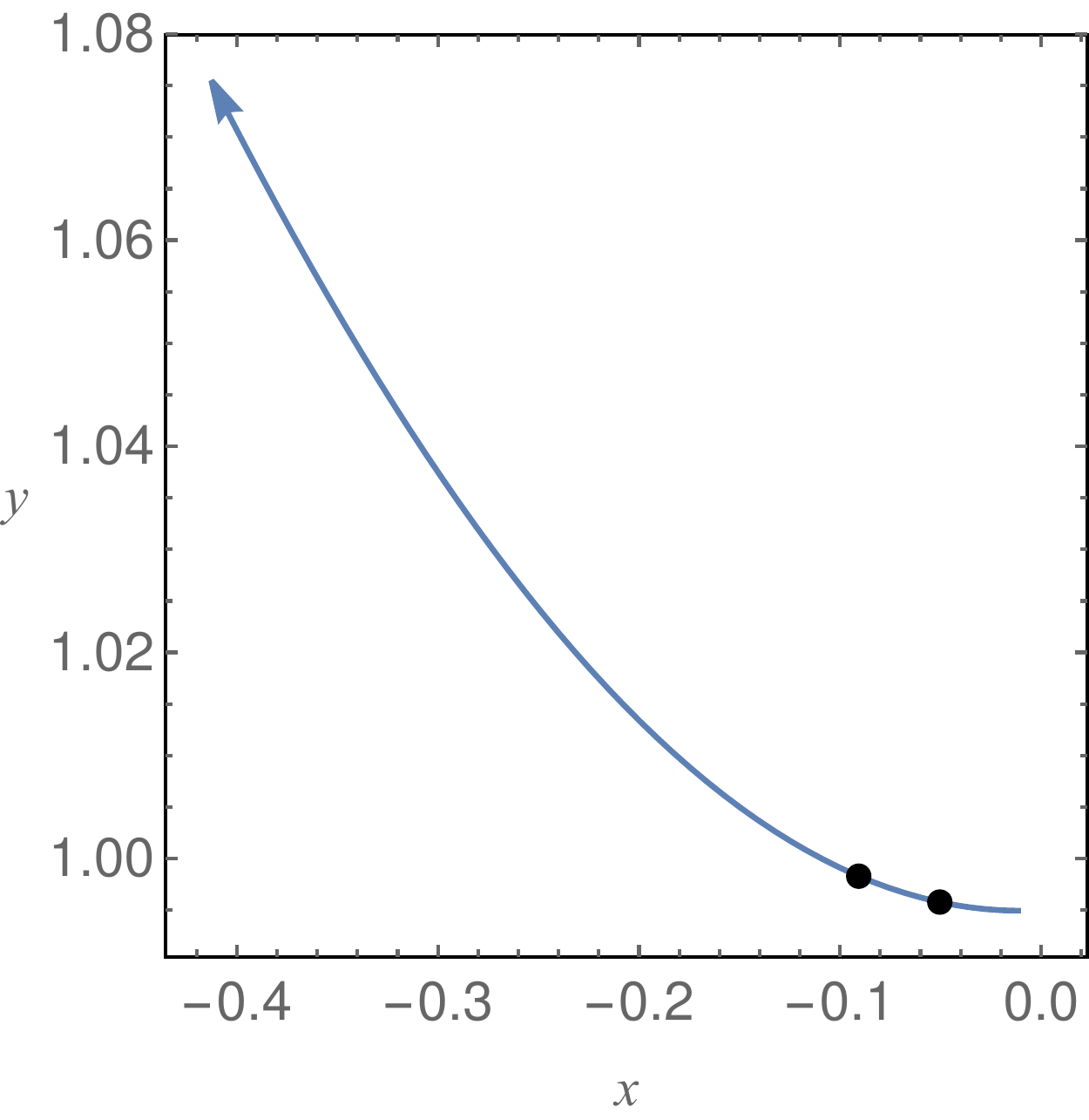}
}
\quad
\subfigure[The variation of the node point $(x_{(A2)}\,,\,y_{(A2)})$, $\alpha$ is fixed as $\alpha=0.2$ and $\xi_0$ varies from 0 to 0.1.  The critical points when $\{\alpha,\xi_0 \} =\{0.2,0.02 \}$ and $\{\alpha,\xi_0 \} =\{0.2,0.05 \}$ are specially mentioned on the curve.]{
	\includegraphics[width=0.26\linewidth]{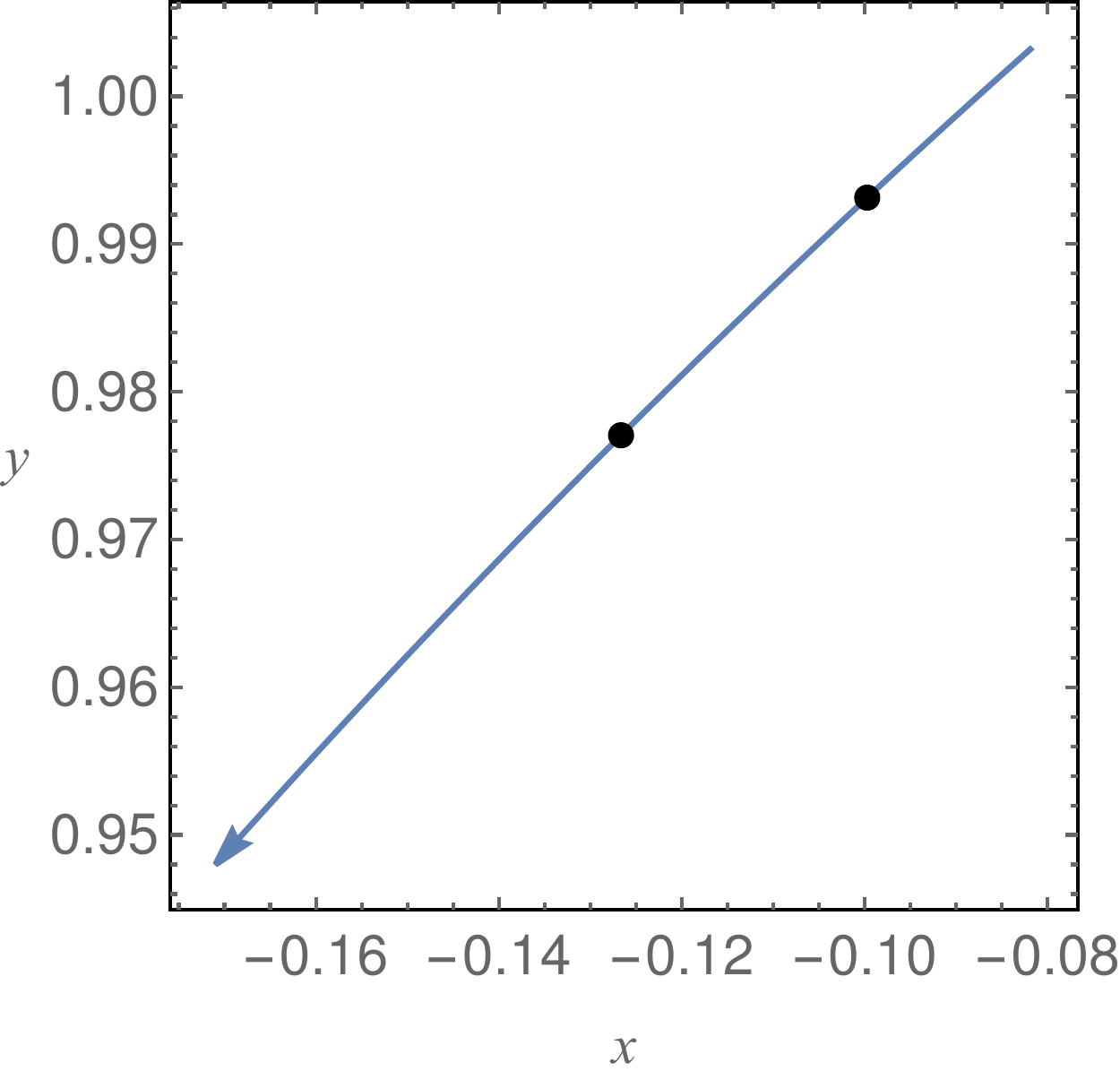}
}
	\label{fig:model1}
	\caption{$\{x,y\}$-phase of Model A. In 
		Fig(a), we have set that $\alpha=0.2$ and $\xi_0=0.05$. In Fig(b) we enlarge the figure and show how the node point varies with $\alpha$, from $\alpha=0$ to $\alpha=1$. The critical point when $\{\alpha,\xi_0 \} =\{0.1,0.01 \}$ and $\{\alpha,\xi_0 \} =\{0.2,0.01 \}$ are specially mentioned on the curve. Similarly, in Fig(c), we set $\alpha=0.2$ with variable $\xi_0$ and  we enlarge the figure to show how the node point varies with $\xi_0$, from $\xi_0=0$ to $\xi_0=0.1$. The critical points when $\{\alpha,\xi_0 \} =\{0.2,0.02 \}$ and $\{\alpha,\xi_0 \} =\{0.2,0.05 \}$ are also specially mentioned on the curve.}
\end{figure}

\section{Autonomous system of  Model B: $\xi_\phi=\xi_0\dot{\phi}$}\label{sec:m2}
Another form for bulk viscosity that we are interested in is $\xi_\phi=\xi_0\dot{\phi}$. The  equations of dynamical system of Model B could be reduced to the following equations:
\begin{eqnarray}
\frac{dx}{dN}&=&-3x-\frac{\sqrt{6}}{2}\alpha y^2+\frac{3}{2}x\bigg[1-x^2-y^2-\xi_0 x\bigg]-3\xi_0x\,,\\
\frac{dy}{dN}&=&-\frac{\sqrt{6}}{2}\alpha xy+\frac{3}{2}y\bigg[1-x^2-y^2-\xi_0x \bigg]\,.
\end{eqnarray}
Similar to Model A,  the critical points for the dynamical system are listed in  Table.III. 
\begin{table}[h]
\centering
\begin{tabular}{ccc}
\hline
\hline
Case &$\,\,\,\,$  Critical points $(x,y)$$\,\,\,\,$  &Stability  \\ \hline
\\
(B1)&$\big(0\,,\,0\big)$&Saddle\\
\\
(B2)&$\big(x_{(B2)},y_{(B2)}\big)$&Stable\\
\\
\hline
\\
\multicolumn{3}{c}{{  $x_{(B2)}=\frac{1}{4\sqrt{6}\alpha}\bigg[6-2\alpha^2-\sqrt{6}\alpha\xi_0+6\xi_0-\sqrt{\big( 6-2\alpha^2-\sqrt{6}\alpha\xi_0+6\xi_0\big)^2+48\alpha^2}\bigg]$\,.}}\\
\\
\multicolumn{3}{c}{{  $y_{(B2)}=\frac{1}{3}\bigg( 3x^2_{(B2)}+(\sqrt{6}\alpha+3\xi_0) x_{(B2)}-3 \bigg)$\,.}}\\
\hline
\hline

\end{tabular}
	\caption{The fixed points and their stability of autonomous system of Model B.}
\end{table}

In Model B, the node point $(x_{(B2)}\,,\,y_{(B2)})$ is always  stable when $\alpha>0$ and $\xi_0>0$, see details in Appendix B.  Next, we study Model B numerically. The  phase diagrams are  given in Fig.3(a), where we choose the parameter to be $\alpha=0.2$ and $\xi_0=0.01$. The critical point in Fig.3(a) is point $(x_{(B2)},y_{(B2)})$ and the different orbits all tend to this attractor. Similar to Model A, the critical point also describes a late time transition from matter dominated era to phantom dominated one.  And as we increase of the value of $\alpha$ and $\xi_0$ individually,  the attractor   also moves away from the point $(0,1)$ in  different directions, which are shown in Fig.3(b) and Fig.3(c).

\begin{figure}[h]
	\centering
	\subfigure[The phase graph of Model B $\{\alpha\,,\,\xi_0\}=\{0.2\,,\,0.01\}$.]{
		\includegraphics[width=0.25\linewidth]{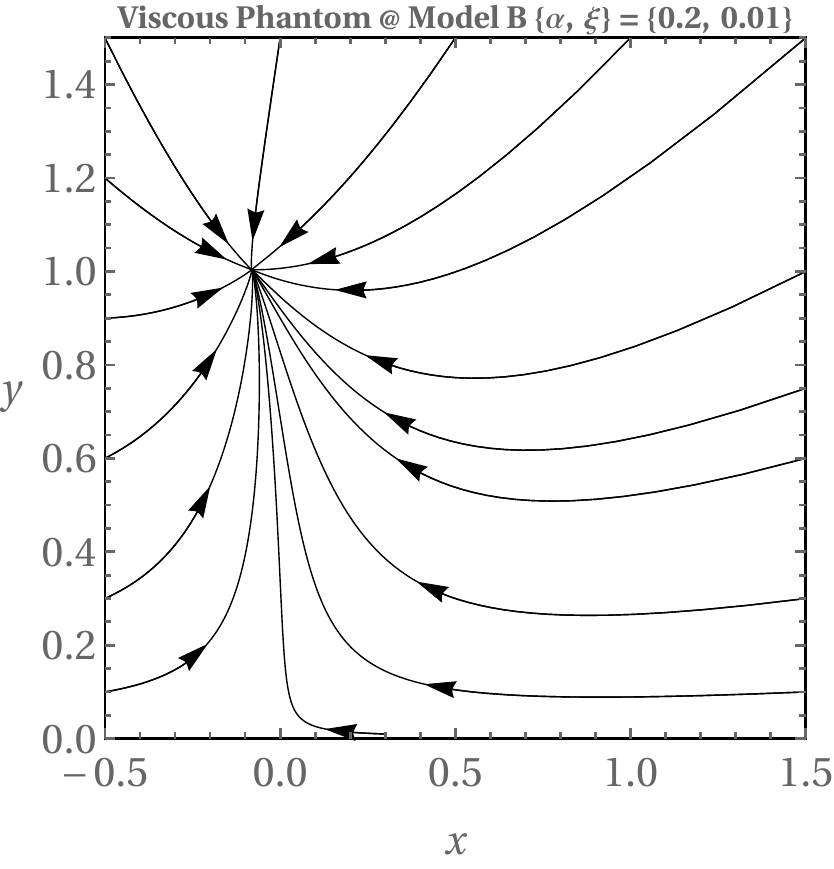}
	}	
	\subfigure[The variation of the node point $(x_{(B2)}\,,\,y_{(B2)})$, $\xi_0$ is fixed as $\xi_0=0.01$ and $\alpha$ varies from 0 to 1. The critical points of $\{\alpha,\xi_0\} =\{0.1,0.01\}$ and $\{\alpha,\xi_0 \} =\{0.2,0.01 \}$ are mentioned on the curve.]{
		\includegraphics[width=0.25\linewidth]{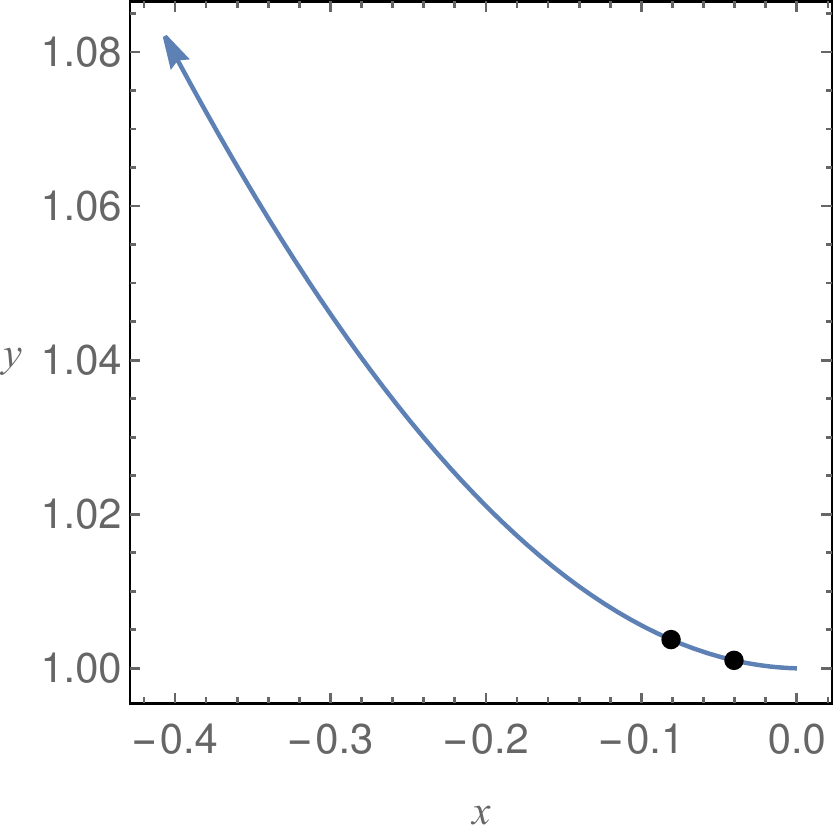}
	}
	\subfigure[The variation of the node point $(x_{(B2)}\,,\,y_{(B2)})$, $\alpha$ is fixed as $\alpha=0.2$ and $\xi_0$ varies from 0 to 0.01. The critical points of $\{\alpha,\xi_0 \} =\{0.2,0.02 \}$ and $\{\alpha,\xi_0 \} =\{0.2,0.05 \}$ are mentioned on the curve.]{
		\includegraphics[width=0.253\linewidth]{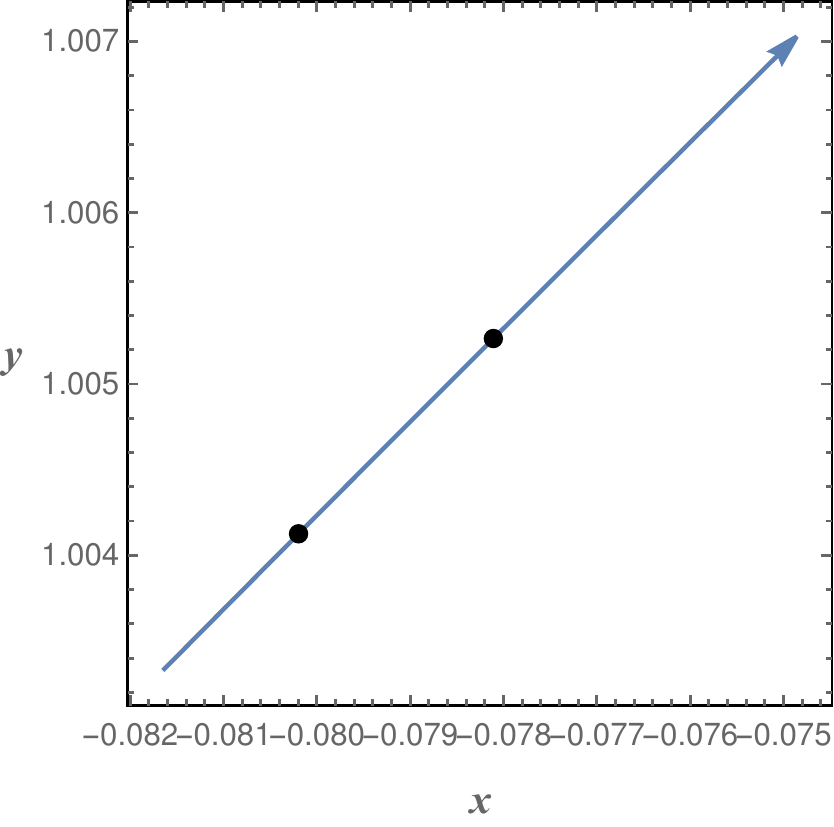}
	}
	\label{fig:model2}
	\caption{$\{x,y\}$-phase of Model B. In 
		Fig(a), we have set that $\alpha=0.2$ and $\xi_0=0.05$.  In Fig(b) we enlarge the figure and show how the node point varies with $\alpha$, from $\alpha=0$ to $\alpha=1$. The critical point when $\{\alpha,\xi_0 \} =\{0.1,0.01 \}$ and $\{\alpha,\xi_0 \} =\{0.2,0.01 \}$ are specially mentioned on the curve. Similarly, in Fig(c), we set $\alpha=0.2$ with variable $\xi_0$ and  we enlarge the figure to show how the node point varies with $\xi_0$, from $\xi_0=0$ to $\xi_0=0.1$. The critical points when $\{\alpha,\xi_0 \} =\{0.2,0.02 \}$ and $\{\alpha,\xi_0 \} =\{0.2,0.05 \}$ are also specially mentioned on the curve.}
\end{figure}

\clearpage

\section{Autonomous system of Model C: Large $\lambda$}\label{sec:m3}

Next, we study a special model  to see the tracking behavior of viscous phantom dark energy. Supposing that $\lambda$ is very large and $\Gamma$ is nearly constant but is not 1, we make the following transformation similar to Ref.\cite{Hao:2003th,Steinhardt:1999nw,Ng:2001hs}. 
\begin{eqnarray}
\epsilon=\frac{1}{\lambda}\,,\,x=\epsilon X\,,\,y=\epsilon Y\,,\,\xi_\phi=\epsilon \xi\,.
\end{eqnarray}
We also suppose that $\xi$ is nearly constant. The terms with $\epsilon$ omitted, the autonomous system can be rewritten in terms of new variables $X$ and $Y$ as:
\begin{eqnarray}
\frac{dX}{dN}&=&-\frac{3}{2}X-\frac{\sqrt{6}}{2}Y^2-3\xi-\sqrt{6}(\Gamma-1)X^2\,,\\
\frac{dY}{dN}&=&\frac{3}{2}Y-\frac{\sqrt{6}}{2}XY-\sqrt{6}(\Gamma-1)XY\,.
\end{eqnarray}
Thus, we may obtain the non-trivial and physically allowed point among the critical points:
\begin{eqnarray}
\big(X_c\,,\,Y_c\big)=\bigg(\frac{3}{2}\Gamma_a\,,\,\frac{\Gamma_a}{\sqrt{2}}\sqrt{\frac{-2\sqrt{6}\xi}{\Gamma_a^2}-\frac{6}{\Gamma_a}+3}\,\,\bigg)\,,
\end{eqnarray}
where $\Gamma_a=\frac{1}{2\Gamma-1}$. At this state, the corresponding energy density parameter of viscous phantom field is given by
\begin{eqnarray}
\Omega_{DE}=-\frac{3}{2}\Gamma_a+\frac{\Gamma_a}{\sqrt{2}}\sqrt{\frac{-2\sqrt{6}\xi}{\Gamma_a^2}-\frac{6}{\Gamma_a}+3}\,,
\end{eqnarray}
and the EoS of viscous phantom energy read as
\begin{eqnarray}
w_{DE}=\frac{-\frac{3}{2}\Gamma_a-\frac{\Gamma_a}{\sqrt{2}}\sqrt{\frac{-2\sqrt{6}\xi}{\Gamma_a^2}-\frac{6}{\Gamma_a}+3}-\xi}{-\frac{3}{2}\Gamma_a+\frac{\Gamma_a}{\sqrt{2}}\sqrt{\frac{-2\sqrt{6}\xi}{\Gamma_a^2}-\frac{6}{\Gamma_a}+3}}\,.
\end{eqnarray}
From the physical requirement $0\leq \Omega_{DE}\leq 1$,  the viable area of $\Gamma$ and $\xi$ is shown in Fig.4.

\begin{figure}[h]
	\centering
	\includegraphics[width=0.3\linewidth]{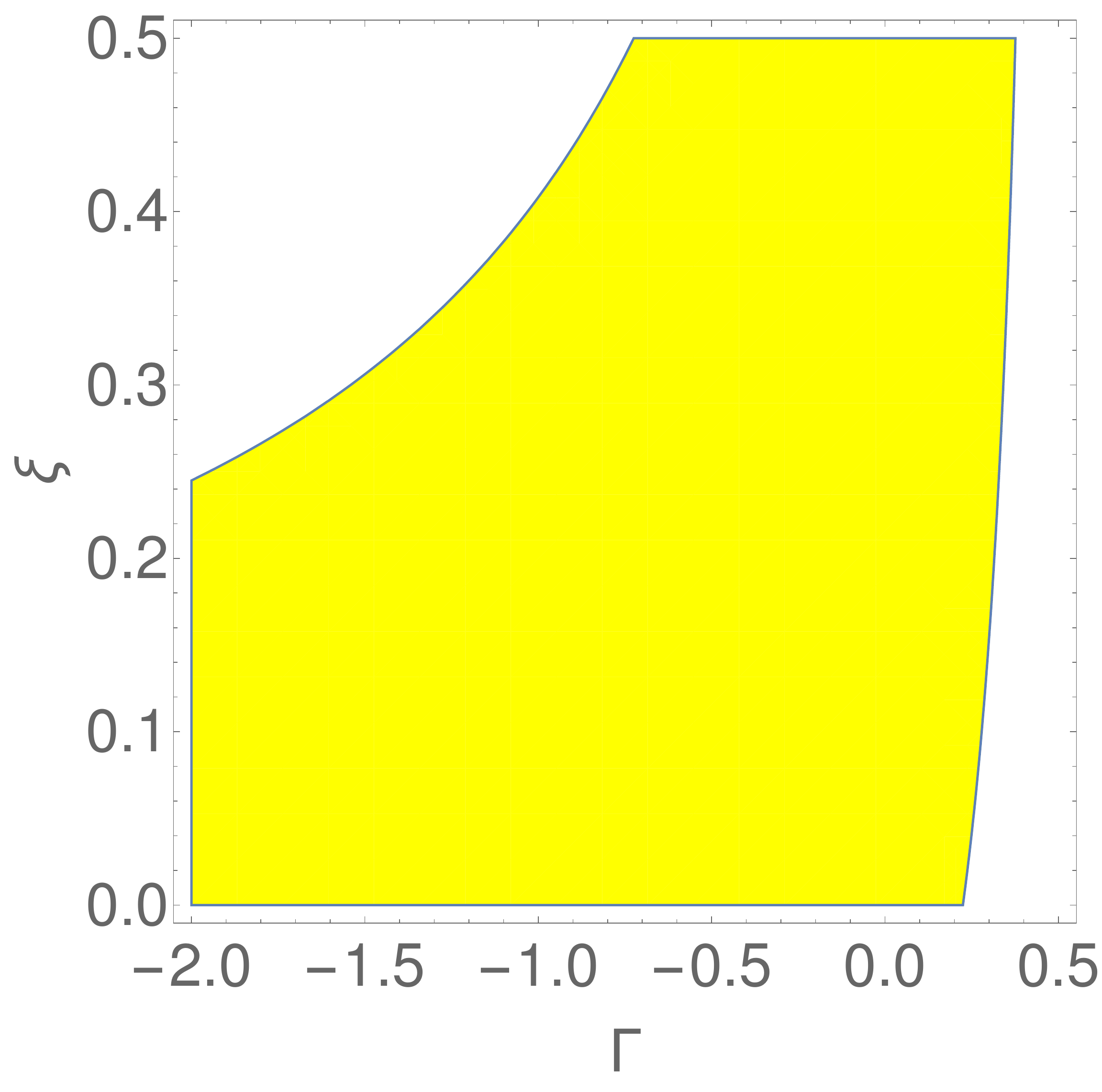}
	\caption{The viable area of $\Gamma$ and $\xi$ in Model C.}
	\label{fig:kexing}
\end{figure}

As remarked before, $\lambda$ is large, thus we can easily find that the critical point $\big(X_c\,,\,Y_c\big)$ leads to a matter dominated Universe. And the density parameter of the viscous phantom field would track with the evolution of dark matter.

Next we will study the stability of the critical point. Similar to Model A and Model B, we have
\begin{eqnarray}
\frac{d (\delta X)}{dN}&=&\bigg[ -\frac{3}{2}-2\sqrt{6}X(\Gamma-1)\bigg]\delta X-\sqrt{6}Y \delta Y \,,\\
\frac{d (\delta Y)}{dN}&=&  -\sqrt{\frac{3}{2}}\frac{Y}{\Gamma_a}\delta X+\frac{1}{2}\bigg( 3-\frac{\sqrt{6}X}{\Gamma_a} \bigg)\delta Y\,.
\end{eqnarray}
The eigenvalues are given by
\begin{eqnarray}
\Lambda_1&=&4\Gamma_a\bigg(-9\Gamma_a+6-\sqrt{3}\sqrt{\Delta_c}\bigg)\,,\\
\Lambda_2&=&4\Gamma_a\bigg(-9\Gamma_a+6+\sqrt{3}\sqrt{\Delta_c}\bigg)\,,
\end{eqnarray}
where
\begin{eqnarray}
\Delta_c=-128 \sqrt{6} \Gamma ^3 \xi +192 \sqrt{6} \Gamma ^2 \xi -84 \Gamma ^2-96 \sqrt{6} \Gamma  \xi +60 \Gamma +16 \sqrt{6} \xi +3\,.
\end{eqnarray}
According to the previous discussion, we found that in the viable area of $(\Gamma,\xi)$ the  eigenvalues are negative. Thus this is a stable point.

Next, we study the dynamical system of Model C numerically.  We can observe that  all the orbits tend to the only non-trivial critical point appearing in the phase space in Fig.5(a), where we choose the parameter to be $\Gamma=-0.8$ and $\xi_0=0.05$. We also study that how the spiral point moves when the parameters are changed, which are also shown in Fig.5(b) and Fig.5(c) .

\begin{figure}[h]
	\centering
	\subfigure[The phase graph of Model C,  $\{\Gamma\,,\,\xi\}=\{-0.80,0.05\}$.]{
		\includegraphics[width=0.24\linewidth]{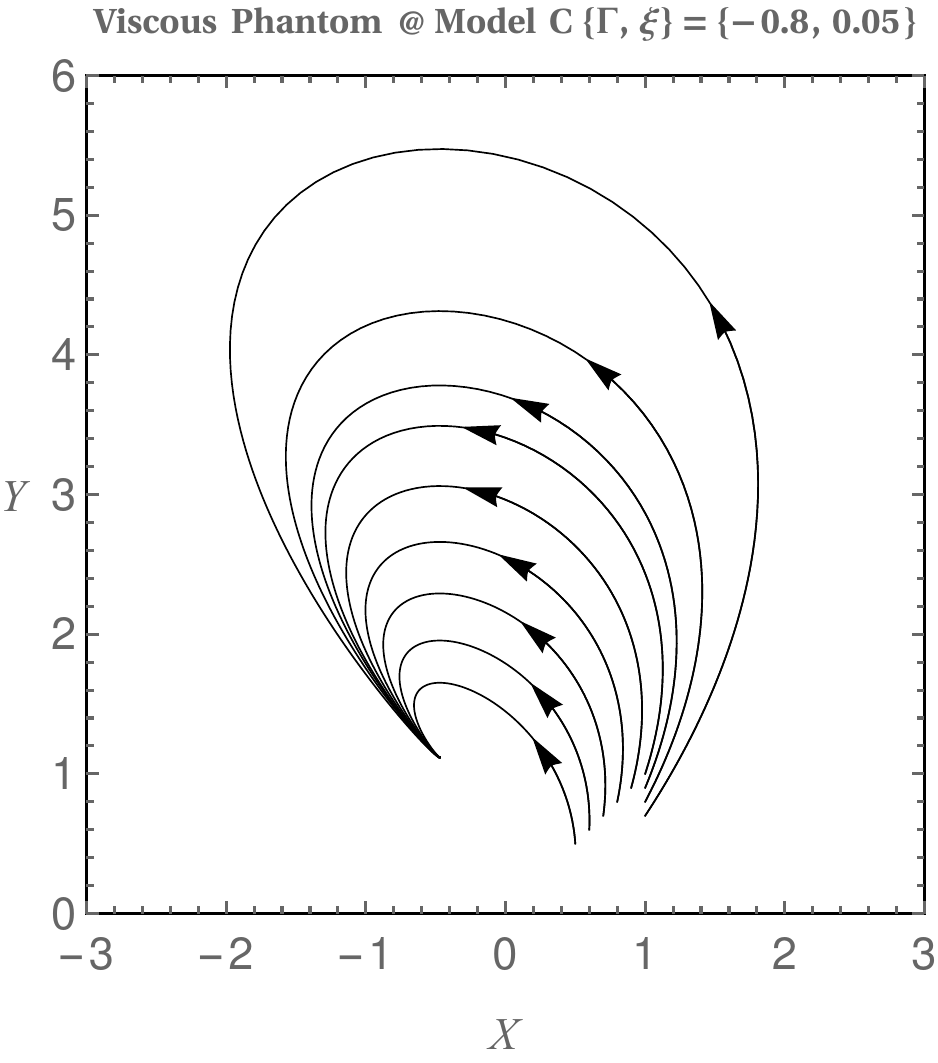}
	}
	\subfigure[The variation of the spiral point $(X_{c}\,,\,Y_{c})$, $\xi$ is fixed as $\xi=0.01$ and $\Gamma$ varies from 0 to -1. The critical points of $\{\Gamma\,,\,\xi\}=\{-0.80,0.01\}$ and $\{\Gamma\,,\,\xi\}=\{-0.80,0.05\}$ are mentioned on the curve.]{
		\includegraphics[width=0.255\linewidth]{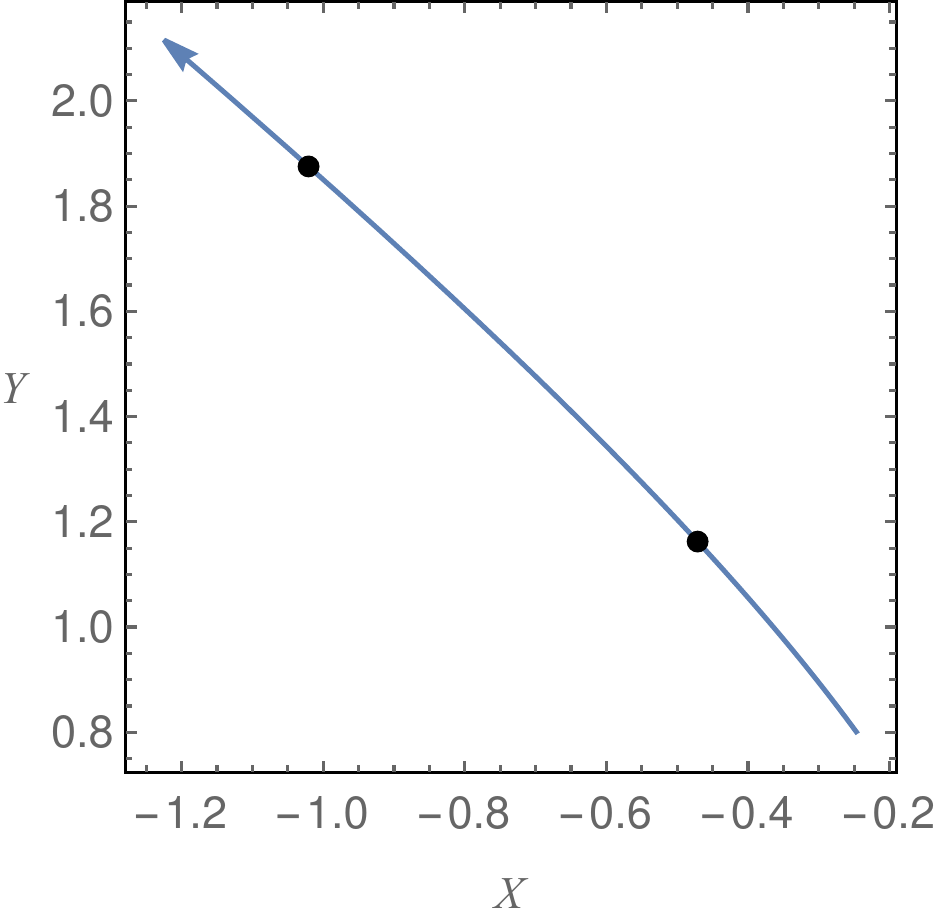}
	}
	\subfigure[The variation of the spiral point $(X_{c}\,,\,Y_{c})$, $\Gamma$ is fixed as $\Gamma=-0.8$ and $\xi$ varies from 0 to 0.1. The critical points of $\{\Gamma\,,\,\xi\}=\{-0.10\,,\,0.05\}$ and $\{\Gamma\,,\,\xi\}=\{-0.80,0.05\}$ are mentioned on the curve.]{
	\includegraphics[width=0.25\linewidth]{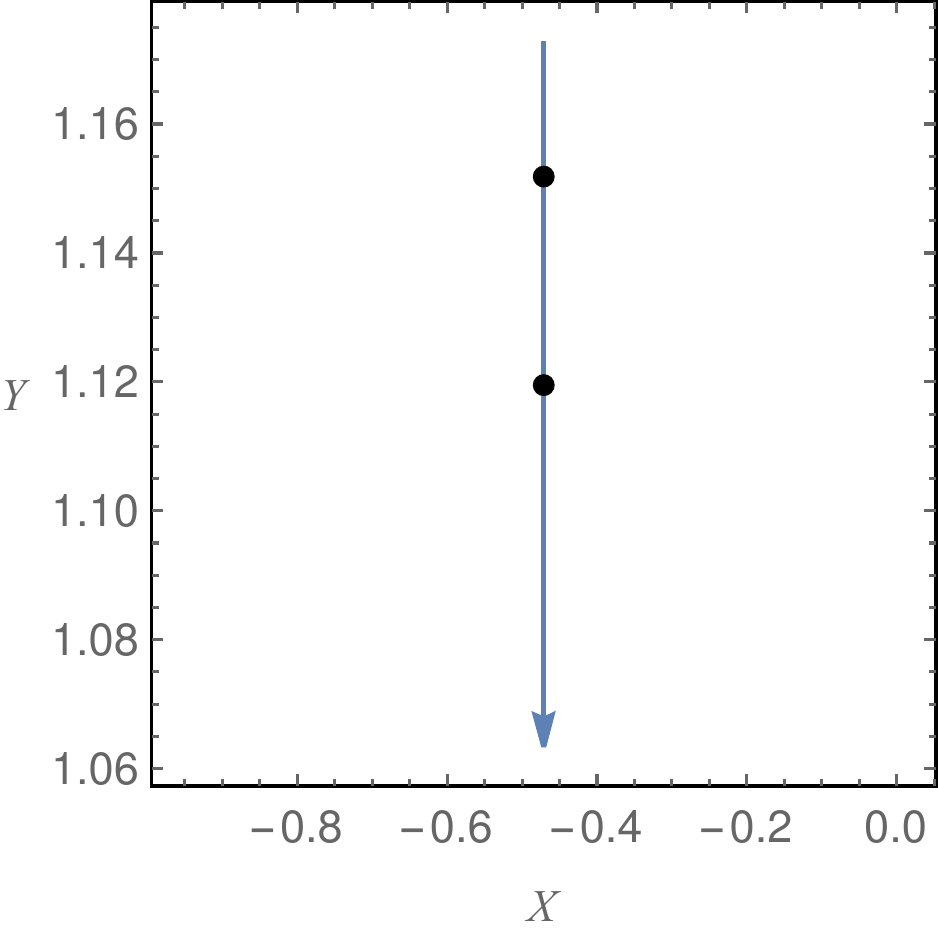}
}
	
	\caption{$\{x,y\}$-phase of Model C. We have set that $\{\Gamma\,,\,\xi\}=\{-0.80\,,\,0.05\}$ in Fig(a). In Fig(b) we set $\xi=0.01$ enlarge the figure and show how the node point varies with $\Gamma$, from $\Gamma=0$ to $\Gamma=-1$. The critical points when $\{\Gamma\,,\,\xi\}=\{-0.10,0.01\}$ and $\{\Gamma\,,\,\xi\}=\{-0.80,0.01\}$ are specially mentioned on the curve. Similarly, in Fig(c), we set $\Gamma=-0.8$ with variable $\xi_0$ and  we enlarge the figure to show how the spiral point varies with $\xi_0$, from $\xi=0$ to $\xi=0.1$. The critical points when $\{\Gamma\,,\,\xi\}=\{-0.80,0.02\}$ and $\{\Gamma\,,\,\xi\}=\{-0.80,0.05\}$ are also specially mentioned on the curve.}
\end{figure}

\clearpage
\section{Statefinder Diagnostic for Viscous Phantom Universe}\label{sec:sts}

As an effective method to distinguish different dark energy models,  statefinder diagnostic is widely applied\cite{Sahni:2002fz,Alam:2003sc,Feng:2008rs,Zimdahl:2003wg,Liu:2007mg,Setare:2006xu,Xi:2017qhj}. The statefinder parameters $\{r,s\}$ are defined as
\begin{eqnarray}
r\equiv\frac{\dddot{a}}{aH^3}\,\,,\,\,s\equiv\frac{r-1}{3(q-1/2)}\,,
\end{eqnarray}
where $q$ is the deceleration parameter
\begin{eqnarray}
q=-\frac{\ddot a}{aH^2}\,.
\end{eqnarray}
Apparently, the statefinder parameters depend on the higher derivative of scale factor. For flat LCDM model, the statefider parameters correspond to a fixed point $\{s,r\}=\{0,1\}$. In order to differentiate among different forms of phantom Universe, we proposed statefinder diagnostic to Model A, Model B and phantom Universe without viscosity(Model P). Using the conservation equations, one can obtain that
\begin{eqnarray}
r  &=&1-\frac32w^\prime_\phi \Omega_\phi+\frac92w_\phi(1+w_\phi)\Omega_\phi\,,\\
s  &=&1-\frac{w^\prime_\phi}{3w_\phi}+w_\phi\,,\\
q  &=&\frac12 (1+3w_\phi \Omega_\phi)\,,
\end{eqnarray}
where $^\prime=\frac{d}{dN}$.

In the following we show the time evolution of statefinder parameters in the case of $\alpha=1$ and $\xi_0=0.02$ in Fig.6(a). We have also plot Model P as a contrast. We can see that the $\{s,r\}$ exists in $r<1$. The plot is for the interval $N\in [0,100]$. Model B and Model P go through the LCDM fixed point M  but Model A behaves differently. The differences between Model B and Model P are that in Model B, the evolution of statefinder parameters firstly exists in the right side of $\{s,r\}=\{0,1\}$, then go back to the LCDM fixed point and next behaves as a climbing up solution as time passes. We also show the trajectories of the statefinder in the $q$–$r$ plane in Fig.6(b).  It is easy to see that in Fig.(a) Model A is almost linear in some stage of evolution, and in Fig.(b) the deceleration parameter changes from nearly one constant to a climbing-up like solution as time passes.

\begin{figure}[h]
	\centering
	\subfigure[Evolving trajectories of the statefinder pairs in the $s$–$r$ plane.]{
		\includegraphics[width=0.4\linewidth]{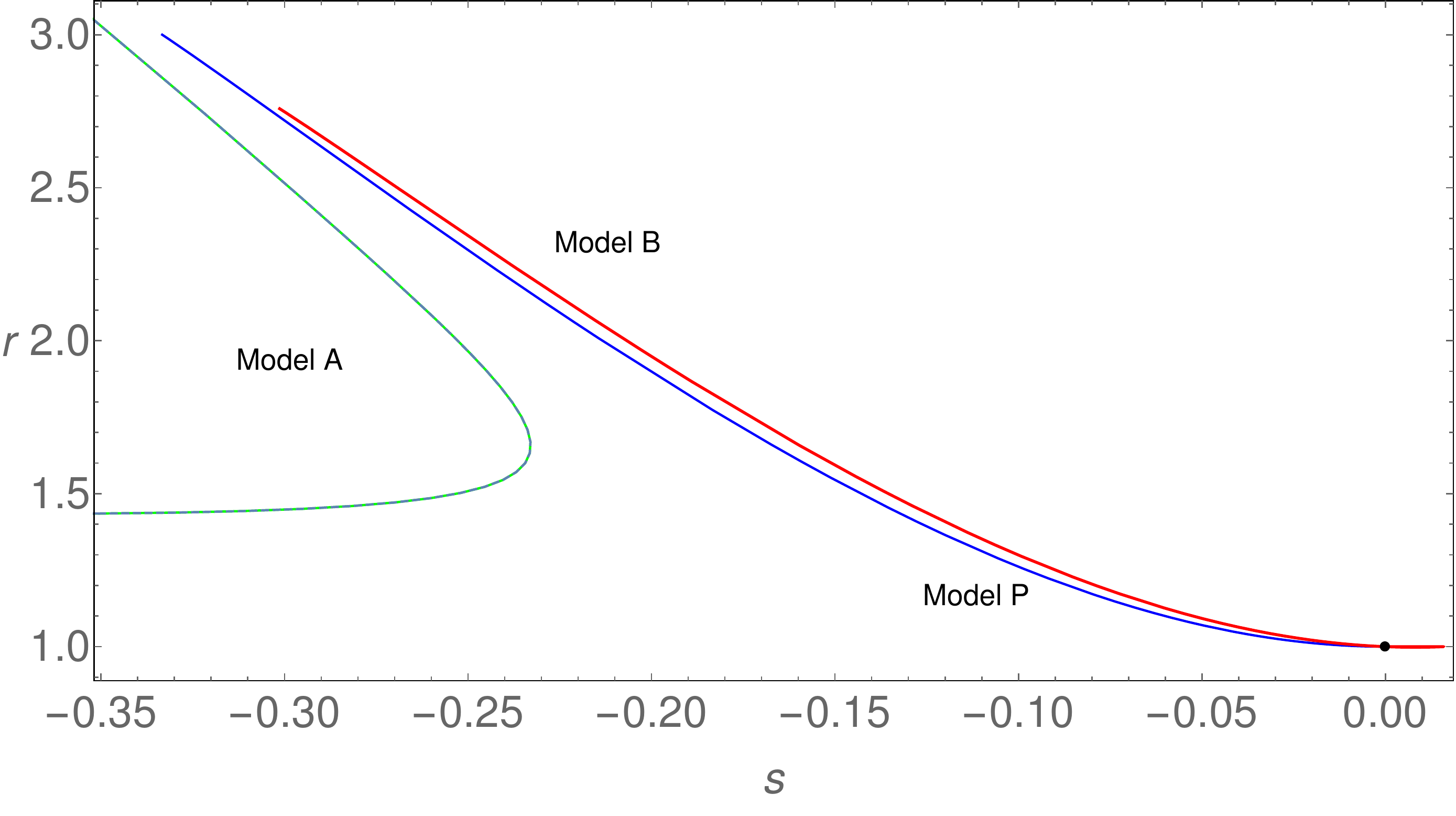}
	}
	\subfigure[Evolving trajectories of the statefinder pairs in  the $q$–$r$ plane.]{
		\includegraphics[width=0.4\linewidth]{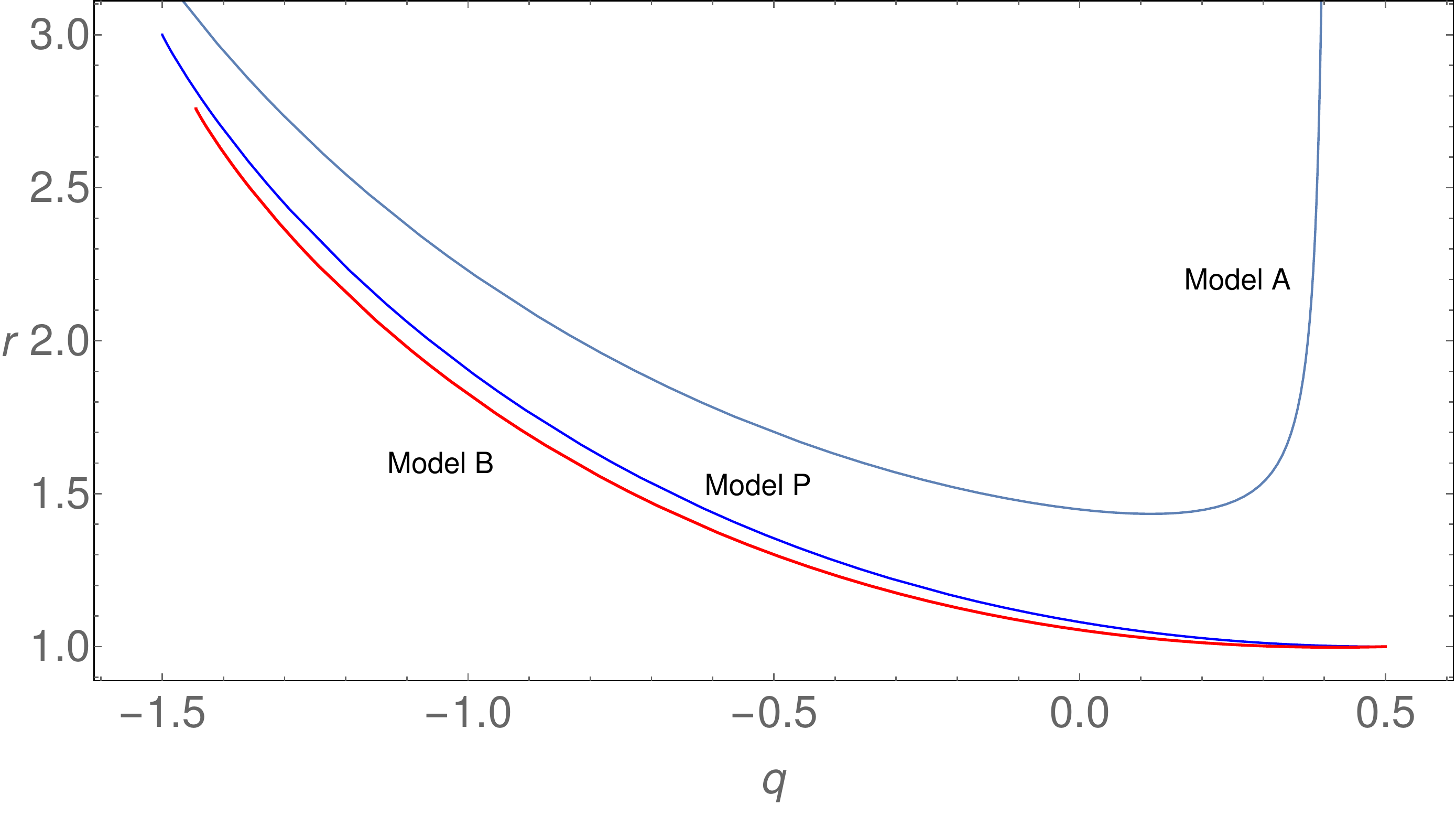}
	}
	\caption{Evolving trajectories of the statefinder in the $s$–$r$ and $q$–$r$ planes for the case of $\alpha=1$ and $\xi_0=0.02$. Model A is
described by the green dashed lines. The red thick lines represent the evolving trajectories of the Model B. Model P is presented as blue lines. The black dot represents LCDM.
}
\end{figure}
We then discuss the statefinder for Model B with different $\xi_0$ to investigate how the viscosity influences the evolution of the Universe. The results are shown in Fig.7, where we choose $\xi_0=0.01$, $\xi_0=0.02$ and $\xi_0=0.03$ as examples. From the statefinder view one can see that with the larger the viscosity is, the evolution of the universe is slowed down  more apparently.  
\begin{figure}[h]
	\centering
	\subfigure[Evolving trajectories of the statefinder pairs of Model B in the $s$–$r$ plane.]{
		\includegraphics[width=0.4\linewidth]{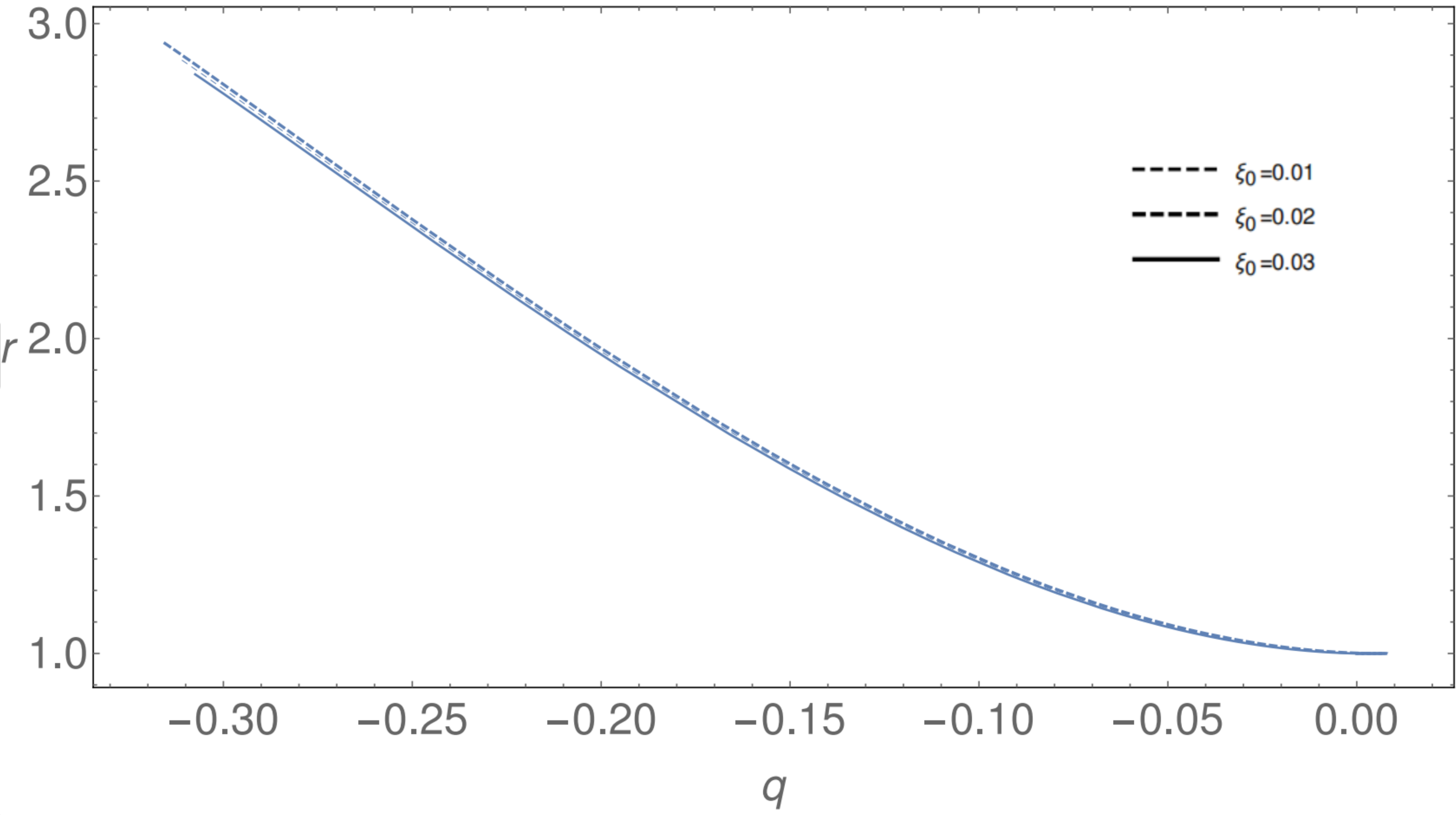}
	}
	\subfigure[Evolving trajectories of the statefinder pairs  of Model B in the $q$–$r$ plane.]{
		\includegraphics[width=0.4\linewidth]{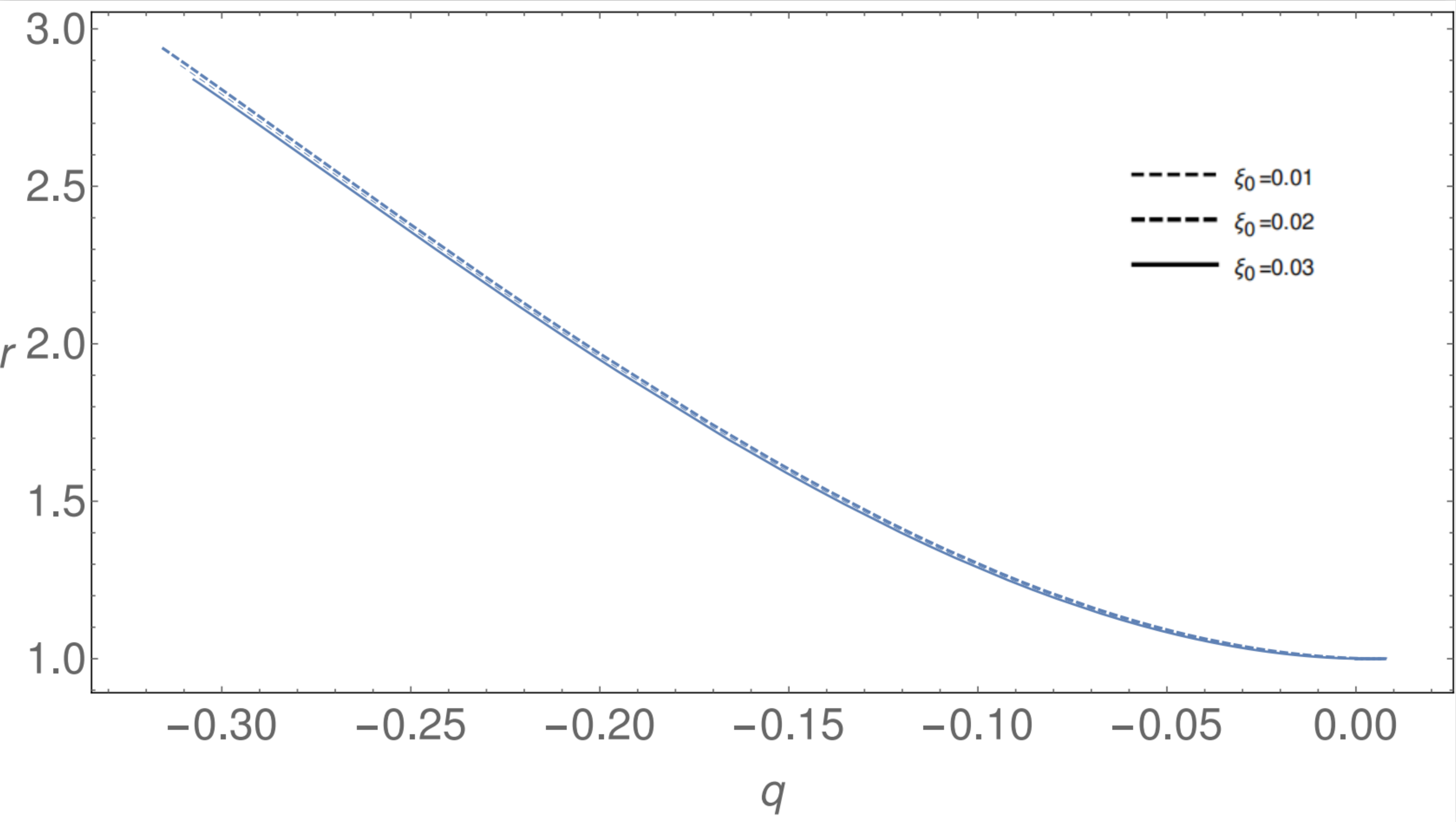}
	}
	\caption{Evolving trajectories of the statefinder in the $s$–$r$ and $q$–$r$ planes for the case of $\alpha=1$ of Model B. The thick dashed lines represent the evolving trajectories of the case of $\xi_0=0.01$, while thin dashed line describes the case of $\xi_0=0.02$ and the case $\xi_0=0.03$ is described by normal lines.}
\end{figure}

\section{Conclusion and Discussion}\label{sec:conclusion}
In this paper, we have investigated the dynamical evolution of three models viscous phantom Universe for different parameters: Model A, $\,\xi_\phi=3\xi_0 H\,$; Model B, $\,\xi_\phi=\xi_0\dot{\phi}\,$; Model C, Large $\,\lambda\,$.  We have shown that in each model, different initial values of the three models will lead to different evolution tracks but the same node point. And for Model A and B, when the viscosity constant $\xi_0$ becomes larger, the node points of the dynamical system will move apparently, which drops a hint that the viscosity does play an important role in the evolution of the Universe. Specially, in Model A, the stability of the critical points also perform constrains to the viscosity constant $\xi_0$ and  $\alpha$ in potential $V=V_0 e^{-\alpha \phi}$.

We also plot the evolving trajectories of the statefinder in the $s$–$r$ and $q$–$r$ planes for the case of $\alpha=1$ and $\xi_0=0.02$. From the statefinder view, we find that Model A describes a big rip Universe, and Model B behaves as a climbing up solution. And we also hope that future high precision observation will be capable of determining these statefinder parameters. 

After all, due to its ability to produce observational predictions cosmology is always a testable theory and we believe that in the future some new experiments with multiple observations and techniques will improve our knowledge of late-time accelerating  expansion. For example, in Ref.\cite{Vagnozzi:2018jhn}, it is demonstrated that quintessence dark energy models can be ruled out in the next 5 years independently of cosmological observations if long-baseline neutrino experiments measure the neutrino mass ordering to be inverted.   And the development of artificial neural network \cite{Cheng:2018nhz} may help in the future study.

\acknowledgments
This work is supported by National Science Foundation of China grant Nos.~11105091 and~11047138, ``Chen Guang" project supported by Shanghai Municipal Education Commission and Shanghai Education Development Foundation Grant No. 12CG51, and Shanghai Natural Science Foundation, China grant No.~10ZR1422000. The authors would like to thank Ping Xi for the useful discussions.

\appendix

\section{The Dynamics of Model A}\label{app:dy1}

In this appendix, we may give a brief discussion for the dynamics of Model A.
\begin{eqnarray}
-3x-\frac{\sqrt{6}}{2}\alpha y^2+\frac{3}{2}x\bigg[1-x^2-y^2-\xi_0\bigg]-3\xi_0&=&0\,,\\
-\frac{\sqrt{6}}{2}\alpha xy+\frac{3}{2}y\bigg[1-x^2-y^2-\xi_0 \bigg]&=&0\,.
\end{eqnarray}
When $y$ is zero, we can obtain that
\begin{eqnarray}
x^3+(1+\xi_0)x+2\xi_0=0\,.
\end{eqnarray}
This is a third order linear equation with standard form that one can easily  solve it by Cardan's formula:
\begin{eqnarray}
-3x+\frac{3}{2}x\bigg[1-x^2-\xi_0\bigg]-3\xi_0=0\,.
\end{eqnarray}
Thus we may obtain the Case (A1) after ignoring the complex solutions.

When $y$ is not zero, we can obtain that
\begin{eqnarray}
0&=&2\sqrt{6}\alpha x^2+2\big(\alpha^2-3\big)x+\sqrt{6}\big(\xi_0-1\big)-6\xi_0\,,\\
y^2&=&1-\frac{\sqrt{6}}{3}\alpha x-\xi_0-x^2\,.
\end{eqnarray}
From Eq.(12), we expect that  $\Omega_{m}>0$.  Thus, the dimensionless variables should satisfy $-x^2+y^2<1$. And this viable critical point is listed in Table I.

For case (A1), $y$ is zero, the perturbation matrix is given by
\begin{eqnarray}
M_1=\left(                 
\begin{array}{cc}   
-\frac{3}{2}\big(1+3x^2+\xi_0\big) &0 \\
\\ 
0 &\frac{1}{2}\big(3-3x^2-9y^2-\sqrt{6}\alpha x-\xi_0\big)\\  
\end{array}
\right)\,.
\end{eqnarray}
The eigenvalues satisfy the equation
\begin{eqnarray}
\lambda^2-\text{Tr}M_1 \lambda+\text{Det}M_1=0\,.
\end{eqnarray}
So we can obtain
\begin{eqnarray}
\lambda_1&=&-\frac{3}{2}\big(1+3x^2+\xi_0\big)\,,\\
\lambda_2&=&\frac{1}{2}\big(3-3x^2-9y^2-\sqrt{6}\alpha x-\xi_0\big)\,.
\end{eqnarray}

When $y$ is not zero, the eigenvalues are given by
\begin{eqnarray}
\lambda_3&=&\frac{1}{4}\bigg[C_1 x+C_2 -\sqrt{D_1x^3+D_2x^2+D_3x+D_4}  \bigg]  \,,\\
\lambda_4&=&\frac{1}{4}\bigg[C_1 x+C_2  +\sqrt{D_1x^3+D_2x^2+D_3x+D_4} \bigg]  \,,
\end{eqnarray}
where
\begin{eqnarray}
C_1&=&3\sqrt{6}\alpha   \,,\\
C_2&=&6\xi_0-12   \,,\\
D_1&=&-96\sqrt{6}\alpha  \,,\\
D_2&=&2\big(-93\alpha^2+72\big)   \,,\\
D_3&=&-4\sqrt{6}\alpha(4\alpha^2+15\xi_0-13)   \,,\\
D_4&=&-48\alpha^2(\xi_0-1)+36\xi_0^2  \,.
\end{eqnarray}
With these results one can discuss the stability of the model.

\section{Dynamic System of Model B}
Similar to Model A, in Model B, the trace and  determinant of the   perturbation matrix are given by
\begin{eqnarray}
\text{Tr} M_2&=&\frac{3}{2}x(\sqrt{6}\alpha+\xi_0)-3(2+\xi_0) \,,\\
\text{Det} M_2&=& 6 \sqrt{6} \alpha x^3+\frac{3}{2} \left(10 \alpha ^2+5 \sqrt{6} \alpha  \xi_0 -6 (\xi_0 +1)\right) x^2\nonumber\\
&&+\left(  \sqrt{6} \alpha ^3+6 \alpha ^2 \xi_0 +3 \sqrt{\frac{3}{2}} \alpha  \big((\xi_0 -2) \xi_0 -6\big)-9 \xi_0  (\xi_0 +1)\right)x-3 \alpha ^2-3 \sqrt{\frac{3}{2}} \alpha  \xi_0 +9 \xi_0 +9\,,
\end{eqnarray}
respectively.
For $\text{Tr} {M_2}<0$, it is easy to see that
\begin{eqnarray}
x_{(B2)}<\frac{2\xi_0+4}{\sqrt{6}\alpha +\xi_0}\,,\nonumber
\end{eqnarray}
when $\alpha$ and $\xi_0$ are both positive. 

So the trace of perturbation matrix is negative for the critical point (B2). And we can read from the matrix that  the determinant of perturbation matrix is positive at the same point. Thus, the eigenvalues for the fixed point are both negative and the fixed point is a stable point.

\begin{equation}
\frac{d\Omega_q}{d\ln a}=-2\Omega_{q}\left(\frac{\sqrt{\Omega_q}}{n}+\frac{1}{H}\frac{dH}{d\ln a}\right),
\end{equation}

\begin{equation}
-\frac{\dot{H}}{H^2}=\frac{3}{2}\left(1-\Omega_q\right)
+\frac{\Omega_q^{3/2}}{n}.
\end{equation}

\begin{equation}
\frac{d\Omega_q}{d\ln a}=
(3-\frac{2}{n}\sqrt{\Omega_q})(1-\Omega_q)\Omega_q.
\end{equation}

\end{document}